# بررسی تاثیر مخارج تحقیق و توسعه بر ارزش افزوده‌ی بخش کشاورزی در ایران

**تانیا خسروی**

استادیار و عضو هیئت علمی، گروه اقتصاد، دانشگاه آزاد اسلامی، واحد کرمانشاه، کرمانشاه، ایران

ta_khosravy@yahoo.com

**سهیل حاتمی‌نیا**

دانشجوی کارشناسی ارشد اقتصاد، دانشگاه آزاد اسلامی، واحد کرمانشاه، کرمانشاه، ایران

hatamysoheyl@gmail.com

**هلیا ضرغامی**

کارشناسی ارشد اقتصاد، دانشگاه یزد، یزد، ایران

Zarghamihelia@gmail.com

**سید فوآد اکبری**

کارشناسی ارشد اقتصاد کشاورزی، دانشگاه فردوسی مشهد، مشهد ، ایران

Foad_a1988@yahoo.com

**چکیده**

در این تحقیق به بررسی تاثیر مخارج تحقیق و توسعه بر ارزش افزوده‌ی بخش کشاورزی در ایران طی دوره های ۱۴۰۱-۱۳۷۹ پرداخته شد. همچنین، با استفاده از روش اقتصادسنجی خود رگرسیون با وقفه های توزیعی (ARDL) ضرایب متغیرهای مستقل تخمین زده شد. نتایج نشان داد که مخارج تحقیق و توسعه در بلند مدت تاثیر مثبت و معنی داری بر روی ارزش افزوده‌ی بخش کشاورزی دارد. در بلند مدت ضریب مخارج تحقیق و توسعه دارای علامت مثبت می‌باشد و از نظر آماری معنی‌دار و برابر ۰.۳۵ است. به عبارت دیگر با افزایش یک درصدی در مخارج تحقیق و توسعه، ارزش افزوده بخش کشاورزی ۰.۳۵ درصد افزایش می‌یابد. همچنین، مطابق انتظار، متغیرهای سرمایه فیزیکی و تعداد شاغلین بخش کشاورزی ( نیروی کار) هم در بلند مدت تاثیر مثبت و معنی داری بر ارزش افزوده‌ی بخش کشاورزی داشتند.

**واژگان کلیدی:** مخارج تحقیق و توسعه، ارزش افزوده‌ی بخش کشاورزی، الگوی ARDL.

**مقدمه**

امروزه رشد اقتصادی مستمر و پایدار از اهداف کلان اقتصادی کشورها است و تحقیق و توسعه یکی از اصلی‌ترین عناصر هر نظام توسعه یافته می‌باشد. تحقیق و توسعه[1] عبارت است از فعالیتی خلاقانه که به صورت نظام‌مند و مستمر جهت افزایش دانش بشری در زمینه‌های مختلف صورت می‌پذیرد و از این دانش برای اختراع یا طرح ایده‌های جدید استفاده می‌گردد(بهلولوند و شمشادی، ۱۴۰۱). در متون اقتصادی تا قبل از سالهای ۱۹۶۰ انباشت سرمایه‌های فیزیکی به عنوان عامل اصلی رشد اقتصادی و یکی از عوامل افزایش بهره‌وری و کارایی بنگاه‌های تولیدی درکشورهای توسعه یافته و درحال توسعه محسوب می‌شد؛ ولی اندیشمندانی چون شولتز[2] و دنیسون[3] نشان دادند که تنها سرمایه فیزیکی نمی‌تواند تفاوت در بهره‌وری و رشد اقتصادی را میان کشورها تبیین کند، بلکه عامل دیگری با عنوان تغییرات توضیح داده نشده باقیمانده وجود خواهد داشت که اثرات آن به مراتب بیشتر از سرمایهٔ فیزیکی است. بنابراین از دههٔ ۱۹۶۰ به بعد، در مورد عوامل مؤثر بر رشد و توسعهٔ اقتصادی کشورها تغییر و تحولات اساسی انجام پذیرفت؛ به نحوی‌که رشد سرمایه انسانی و هزینه‌های تحقیق و توسعه در مدلهای رشد درونزا مورد توجه و تأکید قرار گرفت(بهلولوند و شمشادی، ۱۴۰۱).

تحقیقات کشاورزی یکی از مهمترین عوامل تعیین‌کننده توسعه کشاورزی است، به طوری که در کشورهای توسعه یافته، اعتقاد کلی بر این است که تکنولوژی جدید حاصل تحقیقات دانشمندان در مؤسسات تحقیقات دولتی و خصوصی می‌باشد. هزینه‌های تحقیق و توسعه محصول خود را در شکل تکنولوژی، ابداع و تغییرات فنی وارد تابع تولید می‌نماید. وارد شدن این متغیر در تابع تولید و مدلهای رشد اقتصادی ضمن اینکه در ساختارهای اقتصادی، اجتماعی و فنی جامعه اثر می‌گذارد و آنها را متحول می‌نماید، در بهره‌وری نهاده‌های تولید نیز مؤثر است. تحقیقات کشاورزی، فناوری‌های جدید و بهبود یافته‌ای برای نهاده‌ها و روشهای تولید فراهم می‌کند. با تحقیقات کشاورزی، بهره‌وری منابع افزایش می‌یابد و نهاده‌های جدید با بهره‌وری بالاتر تولید می‌شود. همچنین روشهای نوین برای تولید محصولات غذایی و پتانسیل‌هایی جهت افزایش تولید کشاورزی و کاهش فشار روی منابع طبیعی ایجاد می‌شود. از طرفی تحقیقات کشاورزی می‌تواند باعث وضع سیاستهای جدید و بهبود یافته و یا تغییرات نهادی شود. به عبارت دیگر، این نوع تحقیقات میتواند به کاهش هزینه‌ها کمک کند؛ زیرا افزایش بهره‌وری حاصل از انجام دادن تحقیقات کشاورزی باعث انتقال توابع تولید کشاورزی به سمت بالا می‌شود(مهرابی بشرآبادی و جاودان، ۱۳۹۰).حسب آنچه که بیان گردید در مطالعه حاضر به بررسی تاثیر مخارج تحقیق و توسعه بر ارزش افزوده‌ی بخش کشاورزی در ایران پرداخته خواهد شد.

**پیشینه تحقیق**

**مطالعات داخلی**

مهرابی بشرآبادی و جاودان (۱۳۹۰)، در پژوهشی با عنوان (( تأثیر تحقیق و توسعه بر رشد و بهره‌وری در بخش کشاورزی ایران)) ا استفاده از داده‌های مربوط به سالهای ۱۳۸۶-۱۳۵۳ و الگوی خودتوضیح با وقفه‌های گسترده، به این نتیجه رسیدند که در کوتاه‌مدت و بلندمدت، مخارج تحقیق و توسعه، تأثیر مثبت و معنی‌داری بر رشد و بهره‌وری کل عوامل تولید در بخش کشاورزی ایران دارد. لذا سرمایه‌گذاری در تحقیق و توسعه می‌تواند به عنوان یک منبع اصلی برای رشد بیشتر بخش کشاورزی مدنظر قرار گیرد.

پورعلی مقدم و همکاران(۱۳۹۹)، در پژوهشی با عنوان ((بررسی تأثیر هزینه‌های تحقیق و توسعه، حاکمیت قانون و خشکسالی بر نرخ رشد بخش کشاورزی در ایران)) از اطلاعات دوره زمانی ۹۶-۱۳۸۱و روش شبکه علّی بیزین, نشان دادند که افزایش نرخ

---

[1]. Research and Development: R&D
[2]. Schultz.
[3]. Denison.

رشد هزینه های تحقیق و توسعه، افزایش شاخص حاکمیت قانون و قرار گرفتن کشور ایران در شرایط ترسالی، موجب افزایش بهره وری آب در بخش کشاورزی و رشد این بخش شده است. افزایش هزینه‌های تحقیق و توسعه بهترین عملکرد را در راستای بهبود شاخص خشکسالی نشان داد. به طوری که با قرار دادن احتمال هزینه های تحقیق و توسعه در وضعیت زیاد (۱۰۰ درصد)، بیشترین احتمال شاخص خشکسالی در وضعیت بالا با احتمال ۲۴/۲ درصد قرار گرفت. سناریوی شاخص حاکمیت قانون بهترین عملکرد را در کاهش مصرف آب در بخش کشاورزی نشان داد. به طوری که با قرار دادن احتمال شاخص حاکمیت قانون در وضعیت زیاد، بیشترین احتمال رشد مصرف آب در بخش کشاورزی، در وضعیت پایین با احتمال ۴۹/۵ درصد قرار گرفت. لذا با افزایش هزینه های تحقیق و توسعه و هدفمند کردن آن ها و همچنین اعمال قوانین و سیاست های لازم در زمینه های مختلف از جمله منابع آب می توان به افزایش بهره وری آب در بخش کشاورزی و رشد این بخش کمک کرد.

بهلولوند و شمشادی(۱۴۰۱)، در پژوهشی با عنوان (( اثر مخارج تحقیق و توسعه بر رشد اقتصادی کشورها با استفاده از روشهای PMG و MG (مطالعه موردی ایران و شرکای تجاری منتخب) )) با استفاده از روش میانگین گروهی تلفیقی (PMG) به تحلیل و بررسی اثر مخارج تحقیق و توسعه بر رشد اقتصادی ایران و شرکای منتخب تجاری آن، طی سال‌های (۲۰۱۸- ۲۰۰۱) پرداختند. نتایج نشان داد که در بلندمدت رشد مخارج تحقیق و توسعه با ضریب ۰/۴۶ اثر مثبت و معناداری بر رشد اقتصادی کشورها دارد. همچنین از متغیرهای رشد موجودی سرمایه فیزیکی و نیروی کار شاغل در مدل استفاده شده است و نتایج بیانگر وجود رابطه علّی بلندمدت از سوی متغیرهای مذکور بر رشد اقتصادی است. نتایج کوتاه‌مدت حاصل از روش PMG برای کشور ایران بیانگر آن است که متغیرهای رشد هزینه‌های تحقیق و توسعه، موجودی سرمایه فیزیکی و نیروی کار شاغل دارای اثرات مثبت و معناداری بر رشد اقتصادی ایران در دوره بررسی هستند؛ ازاین‌رو اتخاذ سیاست توسعه اعتبارات به بخش تحقیق و توسعه در کل زیرمجموعه‌های اقتصادی کشور، هم‌زمان با افزایش کیفیت و کمیت عوامل تولید برای دستیابی به رشد اقتصادی مستمر و پایدار ضروری است.

**مطالعات خارجی**

تاری و همکاران[4](۲۰۱۷)، در پژوهشی به بررسی رابطه تحقیق و توسعه و رشد اقتصادی در کشور ترکیه بین سال‌های ۱۹۹۰- ۲۰۱۴ و با استفاده از روش خودرگرسیونی با وقفه‌های توزیعی پرداخته‌اند. نتایج به دست آمده بیانگر آن است که هزینه‌های تحقیق و توسعه دارای اثر مثبت و معناداری بر رشد اقتصادی کشور ترکیه در کوتاه مدت و بلندمدت است. براین اساس سرمایه گذاری در بخش تحقیق و توسعه جهت دستیابی به پایداری رشد اقتصادی در ترکیه مهم ارزیابی شده است.

تسورایی[5](۲۰۱۷)، در مطالعه‌ای به بررسی نقش تحقیق و توسعه بر رشد اقتصادی کشور مجارستان طی سال‌های ۱۹۹۶-۲۰۱۳ پرداخت. نرخ رشد سالانه تولید ناخالص داخلی به عنوان متغیر وابسته وهزینه‌های تحقیق و توسعه به عنوان متغیر مستقل و معیار نوآوری استفاده شد. نتایج به دست آمده از آزمون همگرایی یوهانسن و جوسیلیوس بیانگر آن بود که بین تحقیق و توسعه و رشد اقتصادی در مجارستان رابطه بلندمدتی وجود دارد. همچنین رشد اقتصادی ناشی از هزینه‌های تحقیق و توسعه در مجارستان درکوتاه مدت و بلندمدت، از رابطه علیت گرنجری تبعیت می‌کند. بنابراین افزایش هزینه‌های تحقیق و توسعه به منظور توسعه و رشد اقتصادی کشور مذکور مهم تلقی شده است.

اکالی و همکاران[6](۲۰۱۵)، در پژوهشی به بررسی اثرات مخارج تحقیق و توسعه بر رشد اقتصادی ۱۹ کشور توسعه‌یافته و در حال توسعه طی سال‌های ۱۹۹۰-۲۰۱۳ پرداختند. به این منظور شاخص‌های تولید ناخالص داخلی سرانه و مخارج تحقیق و توسعه با استفاده از روش اثرات ثابت و تصادفی مورد بررسی قرار گرفته است. بر اساس نتایج به دست آمده، اثر مخارج تحقیق و توسعه بر رشد اقتصادی کشورهای مورد مطالعه مثبت و معنادار بوده است؛ بنابراین افزایش مخارج تحقیق و توسعه به مثابه

---

[4]. Tari et al.

[5]. Tsaurai et al.

[6]. Akcali et al.

مبنایی برای خلق نوآوری و رشد و توسعه اقتصادی کشورها مهم ارزیابی شده است.

**روش تحقیق**

این تحقیق از نظر جمع‌آوری آمار و اطلاعات، اسنادی (کتابخانه‌ای) و از نظر هدف کاربردی می باشد. دوره مورد نظر در این تحقیق سال‌های ۱۴۰۱-۱۳۷۹ و قلمرو مکانی ما کشور ایران خواهد بود و با استفاده از روش اقتصادسنجی خود رگرسیون با وقفه های توزیعی (ARDL) ضرایب متغیرهای مستقل تخمین زده می شوند و سپس نتایج حاصله تفسیر می‌شود. علت انتخاب بازه زمانی مذکور آمار قابل دسترس و منتشر شده متغیرهای مدل در سایت بانک جهانی و بانک مرکزی می‌باشد ؛ همچنین جهت تجزیه و تحلیل داده ها از نرم افزار E-views استفاده خواهیم کرد. با انتخاب روش مذکور به دنبال آن هستیم تا اثرات بلند مدت متغیرهای مستقل را بر روی متغیر وابسته توضیح دهیم. در این پژوهش ارزش افزوده بخش کشاورزی متغیر وابسته، هزینه تحقیق و توسعه متغیر مستقل اصلی و موجودی سرمایه و نیروی کار متغیرهای کنترلی می‌باشند.

وجود هم‌انباشتگی بین متغیرهای یک مدل اقتصادی نه تنها نشان‌دهنده‌ی وجود رابطه‌ی تعادلی بلندمدت بین متغیرهای آن مدل است، بلکه به این مفهوم است که می‌توان مدل مورد نظر را به روش OLS برآورد نمود. اما زمانی که حجم نمونه‌ی مورد بررسی کوچک باشد، استفاده از روش OLS باعث خواهد شد که برآوردهای بدست آمده تورش‌دار باشند و در نتیجه استنتاجات آماری که بر اساس آن‌ها انجام می‌گیرد، بی‌اعتبار خواهد بود. برای رفع این مشکل معمولاً از الگوهای پویا که تغییرات کوتاه‌مدت را نیز دربرمی‌گیرند؛ استفاده می‌شود. مدل ARDL پویایی‌های کوتاه‌مدت را در خود گنجانده و در نتیجه موجب می‌شود تا ضرایب الگو با دقت بیشتری برآورد شوند. یک الگوی خود رگرسیونی با وقفه‌های گسترده، به طور کلی به صورت $ARDL(p, q_1, q_2, \ldots, q_n)$ نشان داده می‌شود که اگر $Y_t$ متغیر وابسته و $X_t$ متغیر توضیحی باشد، مدل ARDL به صورت زیر خواهد بود:

$$\alpha(L, P)Y_t = \alpha_0 + \sum_{i=1}^{k} \beta_i(L, q_i) X_{i,t} + u_t \quad (1)$$

که $\alpha_0$ مقدار ثابت، $L$ عملگر وقفه‌ها، $P$ تعداد وقفه‌های به کار رفته برای متغیر وابسته‌ی $Y_t$ و $q$ تعداد وقفه‌های مورد استفاده برای متغیرهای مستقل $X_{i,t}$ است (نوفرستی، ۱۳۷۸).

**یافته‌ها**

با توجه به هدف تحقیق که بررسی تاثیر هزینه تحقیق و توسعه بر ارزش افزوده بخش کشاورزی در ایران با استفاده از مدل خود رگرسیون با وقفه های توزیعی (ARDL) است، تصریح مدل به شکل زیر قابل ارائه می‌باشد :

$$LY = \alpha_0 + \sum_{i=0}^{q_1} \gamma_{1i}(LRD)_{t-i} + \sum_{i=0}^{q_2} \gamma_{2i}(LK)_{t-i} + \sum_{i=0}^{q_3} \gamma_{3i} LL_{t-i} + \sum_{j=1}^{p} \alpha_j Y_{t-j} + \varepsilon_t \quad (2)$$

در این مدل متغیرها به شرح زیر است:

$LY$: لگاریتم ارزش افزوده‌ی بخش کشاورزی
$LRD$: لگاریتم مخارج تحقیق و توسعه
$LK$: لگاریتم سرمایه فیزیکی
$LL$: لگاریتم شاغلین بخش کشاورزی

**بررسی پایایی متغیرها**

قبل از انجام عمل تخمین، به بررسی ایستایی متغیرها با استفاده از روش دیکی فولر تعمیم‌یافته (ADF) پرداخته می‌شود، جهت تخمین مدل به روش ARDL درجه انباشتگی متغیرها نباید بیشتر از یک باشد. جداول ۱ و ۲ نتایج آزمون دیکی فولر تعمیم یافته را نشان می‌دهند. نتایج آزمون ریشه واحد بیانگر آن است که متغیرهای مورد بررسی ما به صورت ترکیبی از I(0) و I(1) هستند. به عبارت دیگر یا در سطح ایستا هستند و یا با یکبار تفاضل‌گیری ایستا شده‌اند. در مدل ARDL مهم این است که متغیرها I(2) و بالاتر نباشند و در حقیقت آزمون‌های ریشه‌ی واحد به این دلیل قبل از به‌کارگیری متغیرها در مدل، بکار گرفته می‌شوند تا اطمینان حاصل شود که متغیرها I(0) و I(1) هستند.

**جدول ۱- خلاصه نتایج حاصل از آزمون ریشه واحد دیکی- فولی تعمیم یافته (ADF) در سطح**

| نتیجه | احتمال | آماره | عرض از مبدا و روند | عرض از مبدا | متغیر |
|---|---|---|---|---|---|
| پایا نیست | ۰.۲۶۶۳ | -۲.۶۴۴۰۷۷ | T | C | LY |
| پایا نیست | ۰.۷۲۹۱ | ۰.۱۸۱۰۶۵ | - | - | LRD |
| پایا نیست | ۰.۰۵۷۹ | -۲.۹۳۰۳۶۳ | - | C | LK |
| پایا | ۰.۰۳۵۰ | -۳.۸۳۲۹۰۱ | T | C | LL |

منبع: یافته‌های تحقیق.

**جدول ۲- خلاصه نتایج حاصل از آزمون ریشه واحد دیکی- فولی تعمیم یافته (ADF) با یکبار تفاضل‌گیری**

| نتیجه | احتمال | آماره | عرض از مبدا و روند | عرض از مبدا | متغیر |
|---|---|---|---|---|---|
| پایا | ۰.۰۰۰۹ | -۴.۹۱۲۲۳۷ | - | C | LY |
| پایا | ۰.۰۰۰۱ | -۴.۵۳۰۷۲۱ | - | - | LRD |
| پایا | ۰.۰۰۰۰ | -۵.۲۲۰۶۰۳ | - | - | LK |

منبع: یافته‌های تحقیق.

طبق نتایج حاصل از آزمون ایستایی به روش دیکی فولر تعمیم‌یافته در جدول ۱ و ۲ مشاهده می‌شود که متغیر لگاریتم تعـداد شاغلین بخش کشاورزی در سطح ایستا می باشد؛ اما، سایر متغیرهای مدل با یک مرتبـه تفاضـل‌گیری ایسـتا می‌شـوند. نتـایج آزمون ریشه واحد بیانگر آن است که متغیرها به صورت ترکیبی از I(0) و I(1) هستند. بـه عبـارت دیگـر یـا در سـطح ایسـتا هستند و یا با یکبار تفاضل‌گیری ایستا شده‌اند. لذا بکارگیری مدل ARDL بلامانع است.

**بررسی وجود رابطه‌ی بلند مدت**

بر اساس آزمون هم‌انباشتگی باند، زمانی می‌توان وجود رابطه‌ی بلندمدت بین متغیرهای مدل را پذیرفت که آماره‌ی F بدست آمده از این روش از حد بالای مقادیر بحرانی ارائه‌شده توسط پسران و همکاران (۲۰۰۱) بزرگ‌تر باشد. جدول ۳ نتایج آزمون هم‌انباشتگی را نشان می‌دهد. همان طور که ملاحظه می‌شود مقدار F محاسبه شده از حد بحرانی بالا بزرگ‌تر است. بنابراین وجود رابطه‌ی بلندمدت تأیید می‌شود.

جدول ۳- نتایج آزمون هم‌انباشتگی باند

| آماره‌ی محاسبه شده $F$ | مقادیر کرانه‌ای برای $K=3$ | | |
|---|---|---|---|
| | سطح معنی داری | حد (1) :$I$ بالا | حد (0) :$I$ پایین |
| ۵.۵۸۹۹۳۲ | ۱۰٪ | ۳/۱ | ۲/۰۱ |
| | ۵٪ | ۳/۶۳ | ۲/۴۵ |
| | ۱٪ | ۴/۸۴ | ۳/۴۲ |

$K$: بیانگر تعداد متغیرهای توضیحی مدل است.

منبع: یافته‌های تحقیق.

**بررسی برقراری فروض کلاسیک**

نتایج حاصل از آزمون‌های تشخیص در جدول ۴ ارائه شده است. اولین آزمون مربوط به خودهمبستگی سریالی است. در آزمون خودهمبستگی فرض صفر عدم وجود خودهمبستگی است، بنابراین اگر فرض صفر پذیرفته شود، وجود خودهمبستگی رد می‌شود. با توجه به نتایج ارائه شده در جدول ۴، احتمال آماره آزمون خودهمبستگی سریالی ۰/۲۰ شده و از ۰/۰۵ بزرگ‌تر است؛ بنابراین، فرض صفر پذیرفته می‌شود و خودهمبستگی وجود ندارد. در آزمون رمزی برای شناسایی شکل تبعی مدل، فرض صفر آن است که مدل به درستی تصریح شده است و اگر فرض صفر پذیرفته شود، شکل تبعی مدل صحیح است. با توجه به نتایج ارائه شده در جدول ۴، احتمال آماره آزمون رمزی ۰/۳۳ شده و از ۰/۰۵ بزرگ‌تر است؛ بنابراین، فرض صفر پذیرفته می‌شود و شکل تبعی مدل صحیح است. در آزمون ناهمسانی واریانس، فرض صفر آن است که واریانس‌ها همسان است و اگر فرض صفر پذیرفته شود، ناهمسانی واریانس وجود ندارد. با توجه به نتایج ارائه شده در جدول ۴، احتمال آماره آزمون ناهمسانی واریانس ۰/۱۱ شده و از ۰/۰۵ بزرگ‌تر است؛ بنابراین، فرض صفر پذیرفته می‌شود و واریانس‌ها همسان است. با توجه نتایج به دست آمده، از آنجایی که احتمال آماره‌های محاسباتی بیشتر از ۵٪ است، می‌توان از برقراری فروض کلاسیک اطمینان حاصل کرد.

جدول۴- نتایج آماره‌های تشخیص مدل

| فروض کلاسیک | آزمون $F$ | | نتیجه |
|---|---|---|---|
| | آماره | احتمال | |
| آزمون خودهمبستگی سریالی | ۱/۷۳ | ۰/۲۰ | عدم خودهمبستگی |
| آزمون رمزی برای شناسایی شکل تبعی مدل | ۰/۹۹ | ۰/۳۳ | صحیح بودن شکل تبعی مدل |
| آزمون ناهمسانی واریانس | ۲/۱۵ | ۰/۱۱ | همسانی واریانس |

منبع: یافته‌های تحقیق.

**تخمین رابطه‌ی بلندمدت**

پس از اطمینان از وجود رابطه‌ی بلندمدت، می‌توان به تخمین رابطه بلندمدت پرداخت. نتایج حاصل از تخمین رابطه بلندمدت به روش ARDL در جدول ۶ ارائه شده است.

جدول ۵- نتایج حاصل از برآورد رابطه بلندمدت

| متغیر | ضریب | آماره t | انحراف معیار | احتمال |
|---|---|---|---|---|
| LRD | 0.358675 | 2.688397 | 0.133415 | 0.0505* |
| LK | 0.501990 | 2.146064 | 0.033911 | 0.0676* |
| LL | 2.381152 | 3.225592 | 0.738206 | 0.0050*** |

منبع: یافته های تحقیق

***، ** و * نشان دهنده ی معنی داری در سطح ۵،۱ و ۱۰ درصد می باشد.

همان‌طور که در جدول ۵ نشان داده شده است، متغیرهای مخارج تحقیق و توسعه، موجودی سرمایه و تعداد شاغلین در بلندمدت تاثیر مثبت و معنی‌داری بر ارزش افزوده‌ی بخش کشاورزی داشته‌اند. به عبارت دیگر، افزایش مخارج تحقیق و توسعه، موجودی سرمایه و تعداد شاغلین بخش کشاورزی هر سه موجب افزایش ارزش افزوده‌ی بخش کشاورزی شده‌اند. ضریب متغیر مخارج تحقیق و توسعه از نظر آماری معنی‌دار و برابر ۰.۳۵ است. به عبارت دیگر با افزایش یک درصدی در مخارج تحقیق و توسعه، ارزش افزوده کشاورزی به میزان ۰.۳۵ درصد افزایش می‌یابد. ضریب متغیر موجودی سرمایه از نظر آماری معنی‌دار و برابر ۰.۵ است. به عبارت دیگر با افزایش یک درصدی در موجودی سرمایه، ارزش افزوده کشاورزی به میزان ۰.۵ درصد افزایش می‌یابد. ضریب متغیر تعداد شاغلین بخش کشاورزی نیز برابر ۲.۳۸ است. به عبارت دیگر، با افزایش یک درصدی در تعداد شاغلین بخش کشاورزی، ارزش افزوده بخش کشاورزی ۲.۳۸ درصد افزایش می‌یابد.

### نتایج حاصل از برآورد مدل تصحیح خطا

آخرین قسمت از تخمین مدل به روش ARDL، به ارائه‌ی مدل تصحیح خطا اختصاص دارد. جمله‌ی تصحیح خطا سرعت تعدیل نسبت به تعادل بلندمدت را نشان می‌دهد. از لحاظ نظری ضریب ECM باید منفی باشد و از لحاظ آماری هم باید معنی‌دار باشد. ضریب تصحیح خطای مدل برابر ۰/۳۴- بوده و از لحاظ آماری نیز معنی‌دار است و این بیانگر این است که در هر دوره حدود ۳۴ درصد از عدم تعادل متغیر وابسته برای رسیدن به تعادل بلندمدت تعدیل می‌شود .

جدول ۶- نتایج حاصل از برآورد مدل ECM

| متغیر | ضریب | آماره‌ی t | انحراف معیار | احتمال |
|---|---|---|---|---|
| ECM (-1) | -۰.۳۴۵۱۵۷ | -۴.۱۱۰۲۰۷ | ۰.۰۸۳۹۷۶ | ۰.۰۰۰۰*** |

منبع: یافته‌های تحقیق.

***، ** و * نشان دهنده ی معنی داری در سطح ۵،۱ و ۱۰ درصد می باشد

### آزمون پایداری ضرایب

آماره‌ی پسماند تجمعی برای آزمون ثبات ساختاری در نمودار ۱ ارائه شده است. چون نمودار در حدفاصل فاصله‌ی اطمینان قرار می‌گیرند، فرضیه‌ی صفر مبنی بر پایداری ضرایب را نمی‌توان رد کرد.

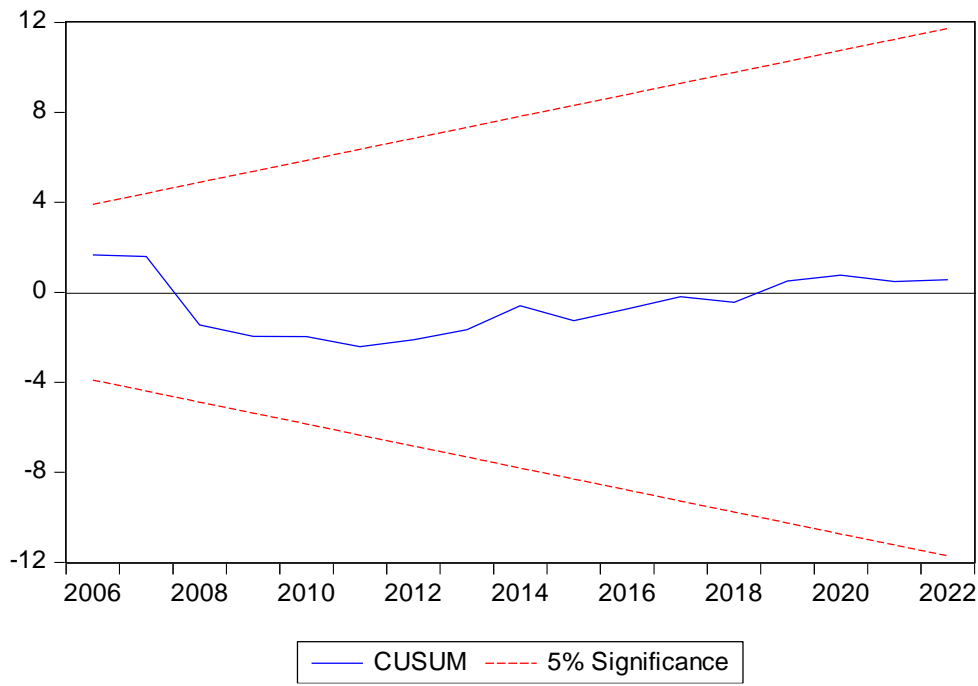

**بحث و نتیجه‌گیری**

در این تحقیق به بررسی تاثیر مخارج تحقیق و توسعه بر ارزش افـزوده‌ی بخـش کشـاورزی طـی دوره هـای ۱۳۷۹-۱۴۰۱ در کشور ایران پرداخته شد. همچنین، با استفاده از روش اقتصادسنجی خود رگرسیون با وقفـه هـای تـوزیعی (ARDL) ضرایب متغیرهای مستقل تخمین زده شد. نتایج نشان داد که مخارج تحقیق و توسعه در بلند مدت تاثیر مثبت و معنـی داری بـر روی ارزش افزوده‌ی بخش کشاورزی دارد. در بلند مدت ضریب مخارج تحقیق و توسعه دارای علامت مثبت می‌باشد و از نظر آمـاری معنی‌دار و برابر ۰.۳۵ است. به عبارت دیگر با افزایش یک درصدی در مخارج تحقیق و توسعه، ارزش افـزوده بخـش کشـاورزی ۰.۳۵ درصد افزایش می‌یابد. همچنین، مطابق انتظار، متغیرهای سرمایه فیزیکی و تعداد شاغلین بخش کشاورزی ( نیروی کـار) هم در بلند مدت تاثیر مثبت و معنی داری بر ارزش افزوده‌ی بخش کشاورزی داشتند. نتـایج ایـن تحقیـق بـا نتـایج مهرابـی بشرآبادی و جاودان (۱۳۹۰)، پورعلی مقدم و همکاران(۱۳۹۹) مبنی برآنکـه مخـارج تحقیـق و توسـعه تـاثیر مثبـت بـر ارزش افزوده‌ی بخش کشاورزی در ایران دارد، همخوانی داشته و نتایج آنها را تایید می‌کند. بنابراین، تحقیقات کشـاورزی، فناوری‌هـای جدید و بهبود یافته‌ای برای نهاده‌ها و روشهای تولید فراهم می‌کند. بـا تحقیقـات کشـاورزی، بهـره‌وری منـابع افـزایش می‌یابـد و نهاده‌هـای جدیـد بـا بهـره‌وری بـالاتر تولیـد می‌شـود. همچنـین روشهای نـوین بـرای تولیـد محصـولات غـذایی و پتانسیل‌هایی جهت افزایش تولید کشاورزی و کاهش فشار روی منابع طبیعی ایجـاد می‌شـود. از طرفـی تحقیقـات کشـاورزی مـی‌توانـد باعـث وضع سیاستهای جدید و بهبود یافته و یا تغییرات نهـادی شـود. بـه عبـارت دیگر، این نوع تحقیقـات میتوانـد به کـاهش هزینـه‌هـا کمـک کند؛ زیرا افزایش بهره‌وری حاصل از انجام دادن تحقیقات کشـاورزی باعـث انتقـال توابـع تولیـد کشاورزی به سـمت بـالا مـی‌شـود

# Investigating the impact of research and development spending on the added value of the agricultural sector in Iran

**Tania Khosravi**

**Assistant Professor and Faculty Member, Department of Economics, Islamic Azad University, Kermanshah Branch, Kermanshah, Iran**

ta_khosravy@yahoo.com

**Soheil Hataminia**

**Master's Student in Economics, Islamic Azad University, Kermanshah Branch, Kermanshah, Iran**

hatamysoheyl@gmail.com

**Helia Zarghami**

**Master of Economics, Yazd University, Yazd, Iran**

Zarghamihelia@gmail.com

**Seyed Fuad Akbari**

**Master of Agricultural Economics, Ferdowsi University of Mashhad, Mashhad, Iran**

Foad_a1988@yahoo.com

**Abstract**

This study examined the impact of research and development expenditures on the value added of the agricultural sector in Iran during the periods 2000-2022. Also, using the econometric method of autoregression with distribution lags (ARDL), the coefficients of the independent variables were estimated. The results showed that research and development expenditures have a positive and significant impact on the value added of the agricultural sector in the long run. In the long run, the coefficient of R&D expenditure has a positive sign and is statistically significant at 0.35. In other words, with a one percent increase in R&D expenditure, the value added of the agricultural sector increases by 0.35 percent. Also, as expected, the variables of physical capital and the number of agricultural workers (labor force) also had a positive and significant effect on the value added of the agricultural sector in the long run.

**Keywords**: Research and Development Expenditures, Agricultural Value Added, ARDL Model.